\documentclass[preprint,12pt,authoryear]{elsarticle}
\usepackage{amssymb}
\usepackage{graphicx}
\usepackage{natbib}
\usepackage[left,modulo]{lineno}

\journal{Environmental Modelling \& Software}

\begin{document}

\begin{frontmatter}

\title{Soil carbon model Yasso07\tnoteref{head1} graphical user interface}
\author[label1,label2]{M. Tuomi\corref{cor1}}\ead{mikko.tuomi@utu.fi}
\author[label3]{J. Rasinm\"aki}\ead{jussi.rasinmaki@simosol.fi}
\author[label1]{A. Repo}
\author[label1]{P. Vanhala}
\author[label1]{J. Liski\fnref{jl}}\ead{jari.liski@ymparisto.fi}
\fntext[label1]{Finnish Environment Institute, Ecosystem Change Unit, Natural Environment Centre, Mechelininkatu 34a, P.O.Box 140, 00251, Helsinki, Finland}
\fntext[label2]{University of Helsinki, Department of Mathematics and Statistics, Gustaf H\"allstr\"omin katu 2b, P.O.Box 68, Helsinki, Finland}
\fntext[label3]{Simosol Oy, Etel\"ainen Asemakatu 2b, 11130, Riihim\"aki, Finland; Telephone: +358 44 040 5859}
\cortext[cor1]{The corresponding author.}
\fntext[jl]{Telephone: +358 40 748 5088; Fax: +358 20 490 2390}
\tnotetext[head1]{Yasso07 modelling project: www.environment.fi/syke/yasso}







\date{29.4.2011}%


\begin{abstract}
In this article, we present a graphical user interface software for the litter decomposition and soil carbon model Yasso07 and an overview of the principles and formulae it is based on. The software can be used to test the model and use it in simple applications. Yasso07 is applicable to upland soils of different ecosystems worldwide, because it has been developed using data covering the global climate conditions and representing various ecosystem types. As input information, Yasso07 requires data on litter input to soil, climate conditions, and land-use change if any. The model predictions are given as probability densities representing the uncertainties in the parameter values of the model and those in the input data -- the user interface calculates these densities using a built-in Monte Carlo simulation.
\end{abstract}

\begin{keyword}
decomposition \sep Monte Carlo simulation \sep software \sep soil carbon \sep statistical modelling \sep uncertainty estimation
\end{keyword}

\end{frontmatter}


\newpage

\begin{linenumbers}

\section*{Software availability}

Software name: Yasso07 User Interface (Yasso07-UI)

Developer: J. Liski, Finnish Environment Institute

Software engineer: J. Rasinm\"aki, Simosol Oy

First available: 2009

Hardware requirements: None

System requirements: Windows, Linux, or OS-X

Software requirements: Python 2.5 on OS-X

Availability: www.environment.fi/syke/yasso

Source code: code.google.com/p/yasso07ui

Programming languages: Fortran 90 (Yasso07), Python (Yasso07-UI)

Software package size: 39Mb (Windows), 34Mb (Linux/\-OS-X)

\section{Introduction}

The decomposition of organic matter (OM) is, in addition to photosynthesis, the other important process that regulates the terrestrial carbon cycle. Decomposition controls carbon dioxide (CO$_{2}$) emissions from soils into the atmosphere and affects the carbon stocks in the soils. Estimates of these emissions and the stock are therefore needed in order to estimate the terrestrial carbon budget. Recently, concern about climate change has increased interest in terrestrial carbon cycle. Terrestrial ecosystems act as remarkable sinks and sources of atmospheric CO$_{2}$ and, consequently, have a significant effect on climate \citep{ipcc2007}.

The carbon cycle of soils has been modelled using a variety of approaches and techniques, and different soil carbon models exist; e.g. Century \citep{parton1987,parton1992}, CoupModel \citep{jansson2004}, Q-model \citep{rolff1999}, ROMUL \citep{chertov2001}, RothC \citep{coleman2005}, DECOMP \citep{wallman2006}, and Yasso07 \citep{liski2005,tuomi2009,tuomi2010}. Due to the high spatial variability of soil cabon stocks \citep[e.g.][]{post1982} and high uncertainty in their changes \citep[e.g.][]{post2001}, and because measuring these variables directly is difficult, laborous, and expensive, soil carbon models are commonly used to estimate these stocks and their changes \citep[e.g.][]{peltoniemi2006,peltoniemi2007,makipaa2008}.

Some of the aforementioned models describe soil carbon cycle at a rather detailed level. For this reason, they also require detailed input information. Such information is not always available, which makes it difficult to apply these models in large geographical scales or at national level. However, the Yasso07 model requires only little easily accessible input information to operate \citep{tuomi2009,tuomi2010}.

Software are already available to operate the CENTURY \citep{parton1987,parton1992} and RothC \citep{coleman2005} soil carbon models but the growing number of practical applications has increased the demand of easily accessible soil carbon models \citep{}. Acknowledging this demand, we constructed a graphical user interface software for the Yasso07 model. This software, hereafter Yasso07-UI, can be used to operate the model and study its performance in estimating the litter decomposition and soil carbon cycle prior to possibly implementing the model into other simulation softwares. Yasso07-UI is also suitable for various practical applications, such as greenhouse gas inventories \citep{climate} and estimation of the effects the removal of forest harvest residues has on carbon balances of boreal forests \citep{repo2011}.

Estimating the modelling uncertainties as reliably as possible has been a topic in several recent studies \citep[e.g.][]{post2008,updegraff2010,ohagan2011}. We acknowledge the importance of accounting for all the sources of uncertainty. In this work these sources are (1) the uncertainties of the model parameters; (2) the uncertainties in the values of the input variables, such as the amount of OM initially in the soil and the litter input to the soil; and (3) the uncertainties in the chemical quality of the litter input. Our software takes all these sources of uncertainty into account by using a built-in Monte Carlo simulation.

\section{Model structure}

\begin{figure}
\centering
\includegraphics[width=\textwidth]{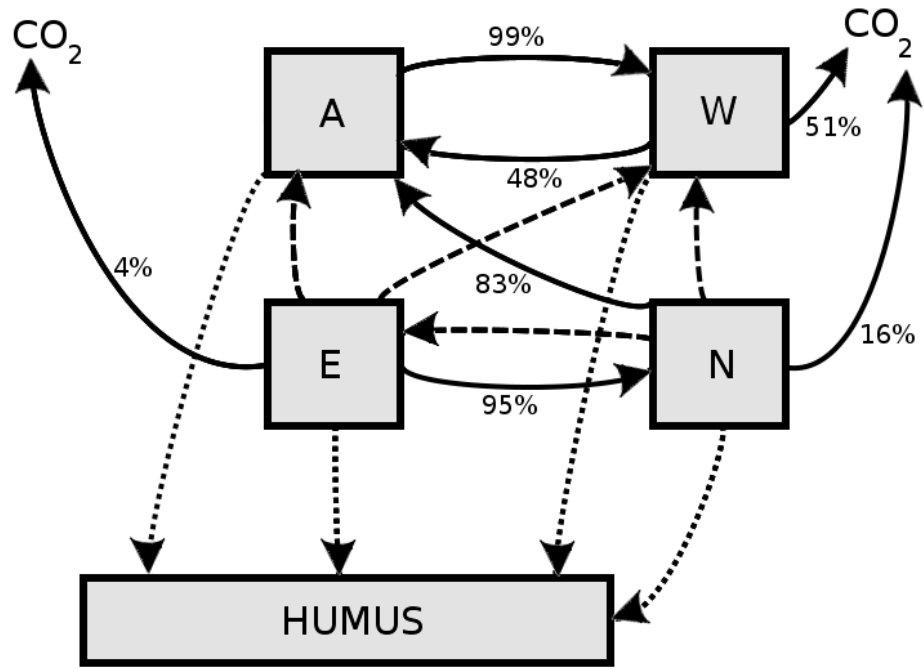}
\caption{Flow diagram of the Yasso07 model and the relative magnitudes of each mass flow between labile compund groups of organic carbon and more calcitrant humus; acid hydrolysables (A), water solubles (W), ethanol solubles (E), and compounds neither soluble nor hydrolysable (N). The carbon flows whose magnitudes differ statistically (95\% confidence) from zero (solid arrows) between and out of the A, W, E, and N fractions (square boxes); the small flows (dotted arrows) into humus (bottom box), each approximately 0.5\%; and the mass flows (dashed arrows) whose MAP estimates were indistinguishable from zero but whose 95\% Bayesian confidence interval was broader than 0.05. See Table \ref{parameters} for details.}
\label{diagram}
\end{figure}

Yasso07 model describes litter decomposition and soil carbon cycle based on the chemical quality of the OM and climatic conditions \citep{tuomi2009}. Decomposition of woody litter depends additionally on the physical size of the litter \citep{tuomi2010}. The model works by dividing fresh OM, e.g. leaf, fine root, and woody litter, into four chemically distinguishable fractions that decompose at their unique rates. These fractions are water solubles (W), ethanol solubles (E), acid hydrolysables (A), and compounds neither soluble nor hydrolysable (N). In addition, there is a humus (H) fraction, assumed to consist of more recalcitrant compounds, that receives a part of the decomposition products of the A, W, E, and N fractions. Carbon flows between the different fractions are shown in Fig. \ref{diagram} \citep[for detailed mathematical formulae describing the Yasso07 model structure, see also][]{tuomi2009}.

With one exception, Yasso07-UI is based on the same structure as was used in \citet{tuomi2009,tuomi2010}. Since there were systematical differences in the mass loss rates between the European and American litter bag measurements used to determine the parameter values of the model earlier \citep{palosuo2005,tuomi2009}, we modelled these differences in a more consistent manner by assuming that a fraction of the OM in the litter bags is transferred out of the bags with varying rates. This process called leaching, can be driven by water flowing through the litter bags. Since the mesh size was different between the European and American litter bags (1.0 mm and 55 $\mu$m, respectively), it is natural that the leaching rates can also be different. The different mesh sizes have an effect also by filtering faunal decomposers in different manners. We modelled the different leaching rates as additional free parameters and replaced the original model (Eq. (1) of \citep{tuomi2009}) by
\begin{equation}\label{modified_model}
  \dot{\mathbf{x}}(t) = \mathbf{A}\mathbf{x}(t) + \mathbf{b} - \omega_{i}\mathbf{I}P_{a}, \mathbf{x}(0) = \mathbf{x}_{0} ,
\end{equation}
where $\mathbf{x}$ is a vector describing the amount of organic carbon in each modelled compartment, $\mathbf{b}$ is the input vector to the system, $\mathbf{A}$ is a decomposition matrix, $\omega_{i}, i=1,2$ are free parameters describing the precipitation induced leaching rates in European ($\omega_{1}$) and American ($\omega_{2}$) litter bags, $\mathbf{I} = (1, 1, 1, 1, 1)^{T}$ is a constant column vector, and $P_{a}$ is the annual precipitation. We selected this model formulation because this effect was not present exclusively in the water soluble compartment W. We tested a model with $\mathbf{I} = (0,1,0,0,0)^{T}$, where the unity corresponds to the compartment containing the water soluble compounds, but this model was found to have much lower posterior probability and it could not fully account for the differences between the European and American litter bags.

The far greater mesh size of the European litter bags likely enables greater leaching rate. This difference may also have an effect on the faunal accessibility to the litter inside the bags, which could increase the mass loss rate in litter bags with greater mesh size slightly \citep{bradford2002,irmler2000}. However, this effect could not be quantified from the available data. Therefore, we simply analysed the data and estimated the parameter probability density functions (PDF's) using Eq. (\ref{modified_model}) but calculate the model predictions in the Yasso07-UI using the original equation because the OM may be leached out of the litter bags but not out of the whole soil system. The parameters $\omega_{i}, i = 1,2$ can in fact be considered as only nuisance parameters that help understanding the process of making the measurements but not the process of decomposition. This improvement had minor effects on the model parameter values -- see Table \ref{parameters} and \citet{tuomi2009}.

\begin{table*}
\centering
\caption{MAP parameter values of Yasso07 and their 95\% confidence limits.\label{parameters}}
\begin{tabular}{lccl}
  \hline \hline
    Parameter & Value & Unit & Interpretation \\
  \hline
    $\alpha_{A}$ & 0.72$\pm$0.09 & a$^{-1}$ & decomposition rate of A \\
    $\alpha_{W}$ & 5.9$\pm$0.8 & a$^{-1}$ & decomposition rate of W \\
    $\alpha_{E}$ & 0.28$_{-0.04}^{+0.07}$ & a$^{-1}$ & decomposition rate of E \\
    $\alpha_{N}$ & 0.031$_{-0.004}^{+0.011}$ & a$^{-1}$ & decomposition rate of N \\
    $p_{1}$ & 0.48$\pm$0.06 & - & relative mass flow, W $\rightarrow$ A \\
    $p_{2}$ & 0.01$^{+0.15}_{-0.01}$ & - & relative mass flow, E $\rightarrow$ A \\
    $p_{3}$ & 0.83$^{+0.16}_{-0.23}$ & - & relative mass flow, N $\rightarrow$ A \\
    $p_{4}$ & 0.99$^{+0.01}_{-0.05}$ & - & relative mass flow, A $\rightarrow$ W \\
    $p_{5}$ & 0.00$^{+0.08}_{-0.00}$ & - & relative mass flow, E $\rightarrow$ W \\
    $p_{6}$ & 0.01$^{+0.20}_{-0.01}$ & - & relative mass flow, N $\rightarrow$ W \\
    $p_{7}$ & 0.00$^{+0.01}_{-0.00}$ & - & relative mass flow, A $\rightarrow$ E \\
    $p_{8}$ & 0.00$^{+0.01}_{-0.00}$ & - & relative mass flow, W $\rightarrow$ E \\
    $p_{9}$ & 0.02$^{+0.23}_{-0.02}$ & - & relative mass flow, N $\rightarrow$ E \\
    $p_{10}$ & 0.00$^{+0.01}_{-0.00}$ & - & relative mass flow, A $\rightarrow$ N \\
    $p_{11}$ & 0.015$\pm$0.015 & - & relative mass flow, W $\rightarrow$ N \\
    $p_{12}$ & 0.95$^{+0.05}_{-0.16}$ & - & relative mass flow, E $\rightarrow$ N \\
    $\omega_{1}$ & -0.151$\pm$0.008 & a$^{-1}$m$^{-1}$ & precipitation induced leaching (Europe) \\
    $\omega_{2}$ & 0.000$^{+0.0}_{-0.002}$ & a$^{-1}$m$^{-1}$ & precipitation induced leaching (Americas) \\
    $\beta_{1}$ & 9.5$\pm$2.0 & $10^{-2}$ $^{\circ}$C$^{-1}$ & temperature dependence \\
    $\beta_{2}$ & -1.4$^{+0.6}_{-0.9}$ & $10^{-3}$ $^{\circ}$C$^{-2}$ & temperature dependence \\
    $\gamma$ & -1.21$\pm$0.14 & m$^{-1}$ & precipitation dependence \\
    $p_{H}$ & 4.5$\pm$0.8 & $10^{-3}$ & mass flow to humus \\
    $\alpha_{H}$ & 1.6$_{-0.2}^{+0.3}$ & $10^{-3}$ a$^{-1}$ & humus decomposition rate\\
    $\phi_{1}$ & -1.71$\pm$0.16 & cm$^{-1}$ & first order size dependence \\
    $\phi_{2}$ & 0.86$\pm$0.10 & cm$^{-2}$ & second order size dependence \\
    $r$ & 0.306$\pm$0.013 & - & size dependence power \\
  \hline \hline
\end{tabular}
\end{table*}

The extension of the model into the woody litter regime is that introduced in \citep{tuomi2010}. Its three parameters describing the effect the physical size of pieces of woody litter have on the decomposition, namely, $\phi_{1}$, $\phi_{2}$, and $r$, are also shown in Table \ref{parameters}.

\section{The coverage of measurements and model predictions}

The data used to construct the Yasso07 model consisted of several individual datasets describing the mass loss rate of leaf, fine root, and woody litter in a variety of climatic conditions \citep{tuomi2009,tuomi2010}. The time and climate coverage of measurements is the range of reliable applicability of the Yasso07 model and the Yasso07-UI introduced here. Despite the fact that the model could be applicable beyond this coverage, the reliability of such applications cannot be known. Therefore, we describe this range briefly.

The climatic range of the measurements of non-woody litter covers more than 90\% of the globally available climatic conditions on land, excluding only the driest deserts and glaciers \citep{tuomi2009}. The litter types included in the data used to develop the model in terms of the initial chemical compositions of litter, cover a wide variety of litters from conifers and broadleaved trees to shrubs and grasses, covering also a variety of different initial nitrogen concentrations \citep{tuomi2009}.The measurements used to develop the model include an extensive data set on decomposition of non-woody litter across Europe, and North and Central America ($N = 9605$), data sets on the decomposition of woody litter in Finland and neighboring regions in Estonia and Russia ($N = 2102$) \citep{tuomi2010}. Therefore, the Yasso07-UI can be applied reliably to a wide variety of litters from different species in almost any climatic conditions.

The time-span of non-woody litter mass-loss measurements covers the first twelve years of decomposition of fresh litter extensively. In longer timescales, of hundreds of years, the litter becomes less important and the decomposition of humus starts to dominate the model predictions because of its low decomposition rate. These timescales are also covered by the available measurements \citep{liski1998,liski2005,tuomi2009} and the model is reasonably trustworthy in predicting the carbon stocks of soils in the global scale as well \citep{tuomi2010b}\footnote{This claim refers to a manuscript in preparation. However, some of these results are accessible through the Yasso07 web site: www.environment.fi/syke/yasso.}. This makes the model trustworthy in predicting the mass loss of non-woody litter in all the timescales.

For woody litter, the measurements used to develop the model include branches and stems ranging from 0.5 to 60 cm in diameter, and the mass loss of these woody biomass components has been followed for 1-70 years since the start of decomposition \citep{tuomi2010}. Hence, the model is reliable in describing decomposition of wood in this timescale.

\section{Yasso07 User Interface}

The Yasso07-UI package contains the Yasso07 model core and the Yasso07-UI software that exploits this model when calculating the predictions. There is also a sample of 100 000 vectors drawn randomly from the posterior probability density of the Yasso07 model parameters included in the package. Using this sample, it is possible to calculate the consequent uncertainty estimates for the model predictions. Because of the availability of this sample, the uncertainty in the measurements, i.e. all the noise and variation not taken into account by the model, is inferenced directly into the predictions of the Yasso07-UI.

\subsection{Input and output information}

The Yasso07-UI works with input information that is easily accessible in practice. This information consists of: 1) the initial state of the soil in terms of the amount of ash free organic carbon in the soil and its chemical quality in terms of AWEN fractions; 2) the climatic conditions as monthly or annual precipitation and mean monthly or annual temperature; and 3) the estimated ash free organic carbon input into the soil in non-woody and woody litter, including the estimated size distribution of the woody litter -- again in terms of the AWEN chemical fractions. The sequential extraction procedures to estimate carbon in the AWEN fractions are described in detail in e.g. \citet{mcclaugherty1985,berg1991a,berg1991b,berg1993,trofymow1998}. A list of chemical compositions of selected litters is provided at the Yasso07 project website (Yasso07-UI manual).

The lack of accurate information on the initial state of the simulated soils is not limiting because given some constant litter input to the soil, the software can be used calculate the long-term stationary state of the soil to be used as the initial state. For modelling purposes, this state is completely defined using the total mass of OM in the soil and its division into the AWEN and H compartments in the model and by making the assumption that the litter input to the soils in terms of amount of carbon is equal to the output.

The litter input to the soil is given to the software by defining the magnitude of the carbon flow into the soil in monthly of annual timescales. It can be given as a constant input or defined as a time-series describing e.g. the annual variations in the input or some long-term trend. The climatic conditions can also be given as some constant values or as time-series with monthly or yearly steps. This enables the user to investigate e.g. the effects climate change or short-scale changes in weather have on the soil carbon stocks and heterotrophic respiration.

The Yasso07-UI presents the modelled estimates for the user in a variety of ways. The results are always available as raw numbers that describe the maximum \emph{a posteriori} (MAP) estimate, the mean, standard deviation, and 95\% confidence intervals, or simply as a sample describing the probability density of the predictions. In addition, the software plots figures showing the amount of OM in each model compartment, the amount of OM in woody and non-woody litters, the total amount of OM, and the estimated respired CO$_{2}$ as a function of time.

\subsection{Uncertainty in the Yasso07-UI}

The uncertainty estimates for model predictions are calculated in two phases. First, a sample is drawn from the joint probability density of the model parameters. When this sample is set large enough (preferrably at least few hundred), it represents the uncertainty in the parameter values of the model caused by the uncertainty and all the unmodelled excess noise in the measurements. In such cases, all the information in the measurements used to construct the model is inferenced to the model predictions as well. The sample size $N = 1$ corresponds to using only the MAP parameter estimates (Table \ref{parameters}) without any uncertainty estimation.

As a next step, the uncertainties in the initial state of the system (i.e. mass of OM and its chemical composition) and in the litter input are taken into account. The initial mass, and the percentage of mass in each compartment, are assumed to be Gaussian random variables. The means and the standard deviations of these values are set by the user and the Yasso07-UI draws random values from these distributions. The litter input is treated similarly.

The uncertainty estimation then proceeds as follows:
\begin{enumerate}
 \item A sample of size $N$ is drawn from the joint posterior density of the model parameters.
 \item For each parameter value $\theta_{i}, i = 1, ..., N$, a random initial state is drawn from the PDF's of the initial state given by the user.
 \item For each parameter value $\theta_{i}$, a random input is drawn from the user defined PDF's of the input for the next time-step.
 \item The state of the system is calculated after the time-step for every $\theta_{i}$ and corresponding initial state and input.
 \item This procedure is repeated by using the new state as the initial state of the next time-step.
 \item The basic estimates describing the probability density of the model predictions; such as mean, standard deviation, and 95\% confidence intervals; are calculated for each time step in the output file.
\end{enumerate}

If the uncertainty of the input information is not known, we recommend the following settings. The uncertainty in the initial mass, namely its standard deviation, should always be set to at least 10\% based on the variations in the available data. This choise would account for the observed variability in the mass remaining measurements of approximately 10 - 15\% of the initial mass \citep{berg1993,gholz2000,tuomi2009}. The uncertainty in the initial chemical composition was found to be approximately 5\% for each chemical component. This is approximately the amount of variation in the chemical composition of litter from a sinle plant species \citep{berg1991a,berg1991b}.

For woody litter, the uncertainties set by the user should always be at least equal to those of non-woody litter. According to the measurements of \citet{palviainen2004,vavrova2009}, the chemical composition of woody litter, in terms of AWEN compounds, does not vary more than that of non-woody litter. Also, the uncertainty in the litter input or initial state mass is not likely to be more than 10\% for woody litter with diameter less than 10 cm. However, for woody litter with diameter greater than this, the uncertainty in the mass should be set to at least 15-20\%. This is roughly the amount of variation in the measurements used to construct the woody litter extension of the Yasso07 model \citep{tuomi2010}. Further, if e.g. the exact size distribution of the woody litter is not known, additional uncertainty should be added to the input and initial state mass estimates to account for all the actual uncertainty in the modelled system.

\section{Conclusions and discussion}

We have introduced a graphical user interface for the litter decomposition and soil carbon model Yasso07. This user interface, Yasso07-UI, can be used to calculate point- and uncertainty estimates for litter decomposition, soil carbon stocks, and CO$_{2}$ emissions from soils in changing conditions with easily available input information. The user interface is applicable in a wide variety of climatic and environmental conditions because it takes into account all the information in the data sets used to build the Yasso07 model \citep{tuomi2009,tuomi2010}.

The Yasso07-UI software can be used as a preliminary tool for assessing whether the Yasso07 model can predict decomposition-related phenomena with respect to some specific set of measurements. These measurements could be e.g. litter mass-loss measurements, soil carbon stock measurements, heterotrophic respiration measurements, or other measurements describing some features of the carbon cycle in soils. If the model appears to be consistent with the data, the Yasso07 model can be easily implemented into different modelling purposes because its source code is freely available, the model has a simple structure, and the input information it requires is commonly readily available.

Because of the nature of the measurements used to build the Yasso07 model, there are cases where it is not known whether the model is applicable or not. These cases include swamps and marshlands, where the limited availability of free oxygen limits the decomposition heavily. This effect is not taken into account by our model. Also, the applicability of our model to modelling carbon stocks of mineral soils has not been tested throughly. Further, the model predictions regarding the decomposition of woody litter remain to be compared to wood decomposition data from temperate and tropical climates. We aim at developing the model further by extending its range of reliable applicability to these environmental and climatic conditions.

\section*{Acknowledgements}

This work was funded by the Maj and Tor Nessling Foundation (project "Soil carbon in Earth System Models") and the Academy of Finland (project 107253).


\end{linenumbers}


\begin{thebibliography}{100}\small
\bibitem[\protect\astroncite{Berg et al.}{1991a}]{berg1991a} Berg, B., Booltink, H., Breymeyer, A., Ewertsson, A., Gallardo, A., Holm, B., Johansson, M.-B., Koivuoja, S., Meentemeyer, V., Nyman, P., Olofsson, J., Pettersson, A.-S., Reurslag, A., Staaf, H., Staaf, I., and Uba, L. 1991a. Data on needle litter decomposition and soil climate as well as site characteristics for some coniferous forest sites, Part I, Site characteristics. Report 41, Swedish University of Agricultural Sciences, Departnent of Ecology and Environmental Research, Uppsala.
\bibitem[\protect\astroncite{Berg et al.}{1991b}]{berg1991b} Berg, B., Booltink, H., Breymeyer, A., Ewertsson, A., Gallardo, A., Holm, B., Johansson, M.-B., Koivuoja, S., Meentemeyer, V., Nyman, P., Olofsson, J., Pettersson, A.-S., Reurslag, A., Staaf, H., Staaf, I., and Uba, L. 1991b. Data on needle litter decomposition and soil climate as well as site characteristics for some coniferous forest sites, Part II, Decomposition data. Report 42, Swedish University of Agricultural Sciences, Departnent of Ecology and Environmental Research, Uppsala.
\bibitem[\protect\astroncite{Berg et al.}{1993}]{berg1993} Berg, B., Berg, M. P., Bottner, P., Box, E., Breymeyer, A., De Anta, R. C., Couteaux, M., M\"alk\"onen, E., McClaugherty, C., Meentemeyer, V., Munoz, F., Piussi, P., Remacle, J., and De Santo, A. V. 1993. Litter mass loss in pine forests of Europe and Eastern United States: some relationships with climate and litter quality. Biogeochemistry, 20, 127-159.
\bibitem[\protect\astroncite{Bradford et al.}{2002}]{bradford2002} Bradford, M. A., Tordoff, G. M., Eggers, T., Jones, T. H., and Newington, J. E. 2002. Microbiota, fauna, and mesh size interactions in litter decomposition. Oikos, 99, 317-323.
\bibitem[\protect\astroncite{Chertov et al.}{2001}]{chertov2001} Chertov, O. G., Komarov, A. S., Nadporozhskaya, M., Bykhovets, S. S., and Zudin, S. L. 2001. ROMUL -- a model of forest soil organic matter dynamics as a substantial tool for forest ecosystem modeling. Ecological Modelling, 138, 289-308.
\bibitem[\protect\astroncite{Coleman and Jenkinsson}{2005}]{coleman2005} Coleman, D. C. and Jenkinson, D. S. 2005. RothC-26.3 A model for turnover of carbon in soil. Model description and Windows users guide. IACR-Rothamsted, Harpenden, U.K., 45 p.
\bibitem[\protect\astroncite{Finland's Fifth National Communication under the United Nations Framework Convention on Climate Change}{2009}]{climate} Finland's Fifth National Communication under the United Nations Framework Convention on Climate Change. 2009. Ministry of the Environment and Statistics Finland, Helsinki. 280 p.
\bibitem[\protect\astroncite{Gholz et al.}{2000}]{gholz2000} Gholz, H. L., Wedin, D. A., Smitherman, S. M., Harmon, M. E., and Parton,  W. J. 2000. Long-term dynamics of pine and hardwood litter in contrasting environments: Toward a global model of decomposition. Global Change Biology, 6, 751-765.
\bibitem[\protect\astroncite{IPCC}{2007}]{ipcc2007} IPCC 2007: Climate Change 2007: The Physical Science Basis. Contribution of Working Group I to the Fourth Assessment Report of the Intergovernmental Panel on Climate Change. Solomon, S., Qin, D., Manning, M., Chen, Z., Marquis, M., Averyt, K. B., Tignor, M., and Miller, H. L. (eds.). Cambridge University Press, Cambridge, United Kingdom and New York, NY, USA, 996 pp.
\bibitem[\protect\astroncite{Irmler}{2000}]{irmler2000} Irmler, U. 2000. Changes in the fauna and its contribution to mass loss and N release during leaf litter decomposition in two deciduous forests. Pedobiologia, 44, 105-118.
\bibitem[\protect\astroncite{Jansson and Karlberg}{2004}]{jansson2004} Jansson, P. and Karlberg, L. 2004. COUP manual. Coupled heat and mass transfer model for soil-plant-atmosphere systems. Stockholm, Sweden, 445 pp.
\bibitem[\protect\astroncite{Liski et al.}{1998}]{liski1998} Liski, J., Ilvesniemi, H., M\"akel\"a, A., and Starr, M. 1998. Model analysis of the effects of soil age, fires and harvesting on the carbon storage of boreal forest soils. European Journal of Soil Science, 49, 407-416.
\bibitem[\protect\astroncite{Liski et al.}{2005}]{liski2005} Liski, J., Palosuo, T., Peltoniemi, M., and Siev\"anen, R. 2005. Carbon and decomposition model Yasso for forest soils. Ecological Modelling, 189, 168-182.
\bibitem[\protect\astroncite{M\"akip\"a\"a et al.}{2008}]{makipaa2008} M\"akip\"a\"a, R., H\"akkinen, M., Muukkonen, P., and Peltoniemi, M. 2008. The costs of monitoring changes in forest soil carbon stocks. Boreal Environmental Research, 13, 120-130.
\bibitem[\protect\astroncite{McClaugherty et al.}{1985}]{mcclaugherty1985} McClaugherty, C. A., Pastor, J., Aber, J. D., and Melillo, J. M. 1985. Forest litter decomposition in relation to soil nitrogen dynamics and litter quality. Ecology, 66, 266-275.
\bibitem[\protect\astroncite{O'Hagan}{2011}]{ohagan2011} O'Hagan, A. 2011. Probabilistic uncertainty specification: Overview, elaboration techniques and their application to a mechanistic model of carbon flux. Environmental Modelling and Software, In press.
\bibitem[\protect\astroncite{Palosuo et al.}{2005}]{palosuo2005} Palosuo, T., Liski, J., Trofymow, J. A., and Titus, B. 2005. Litter decomposition affected by climate and litter quality -- testing the Yasso model with litterbag data from the Canadian Intersite Decomposition Experiment. Ecological Modelling 189, 183-198.
\bibitem[\protect\astroncite{Palviainen et al.}{2004}]{palviainen2004} Palviainen, M., Fin\'er, L., Kurka, A.-M., Mannerkoski, S., Piirainen, S., and Starr, M. 2004. Decomposition and nutrient release from logging residues after clear-cutting of mixed boreal forest, Plant and Soil, 263, 53-67.
\bibitem[\protect\astroncite{Parton et al.}{1987}]{parton1987} Parton, W. J., Schimel, D. S., Colem C. V., and Ojima, D. S. 1987. Analysis of factors controlling soil organic matter levels in Great Plains grasslands. Soil Science Society of America Journal, 51, 1173-1179.
\bibitem[\protect\astroncite{Parton et al.}{1992}]{parton1992} Parton, W. J., McKeown, B., Kirchner, V., and Ojima, D. S. 1992. CENTURY Users Manual. NREL Publication, Colorado State University, Fort Collins, Colorado, U.S.
\bibitem[\protect\astroncite{Peltoniemi et al.}{2006}]{peltoniemi2006} Peltoniemi, M., Palosuo, T., Monni, S., and  M\"akip\"a\"a, R. 2006. Factors affecting the uncertainty of sinks and stocks of carbon in Finnish forests soils and vegetation. Forest Ecology and Management, 232, 75-85.
\bibitem[\protect\astroncite{Peltoniemi et al.}{2007}]{peltoniemi2007} Peltoniemi, M., Th\"urig, E., Ogle, S., Palosuo, T., Schrumpf, M., Wutzler, T., Butterbach-Bahl, K., Chertov, O., Komarov, A., Mikhailov, A., G\"arden\"as, A., Perry, C., Liski, J., Smith, P. and M\"akip\"a\"a, R. 2007. Models in country scale carbon accounting of forest soils. Silva Fennica, 41, 575-602.
\bibitem[\protect\astroncite{Post et al.}{2008}]{post2008} Post, J., Hattermann, F. F., Krysanova, V., and Suckow, F. 2008. Parameter and input data uncertainty estimation for the assessment of long-term soil organic carbon dynamics. Environmental Modelling and Software, 23, 125-138.
\bibitem[\protect\astroncite{Post et al.}{1982}]{post1982} Post, W. M., Pastor, J., Zinke, P. J., and Stangenberger, A. G. 1982. Soil carbon pools and world life zones. Nature, 317, 613-616.
\bibitem[\protect\astroncite{Post et al.}{2001}]{post2001} Post, W. M., Izaurralde, R. C., Mann, L. K., and Bliss, N. 2001. Monitoring and verifying changes of organic carbon in soil. Climatic Change 51, 73–99.
\bibitem[\protect\astroncite{Repo et al.}{2011}]{repo2011} Repo, A., Tuomi, M., and Liski, J. 2011. Indirect carbon dioxide emissions from producing bioenergy from forest harvest residues. Global Change Biology Bioenergy, 3, 107-115. 
\bibitem[\protect\astroncite{Rolff and \AA{}gren}{1999}]{rolff1999} Rolff, C. and \AA{}gren, G. I., 1999. Predicting effects of different harvesting intensities with a model of nitrogen limited forest growth. Ecological Modelling, 118, 193-211.
\bibitem[\protect\astroncite{Trofymow et al.}{1998}]{trofymow1998} Trofymow, J. A. and the CIDET Working Group, 1998. The Canadian Intersite Decomposition ExperimenT (CIDET): project and site establishment report. Information report BC-X-378, Pacific Forestry Centre, Victoria, Canada.
\bibitem[\protect\astroncite{Tuomi et al.}{2009}]{tuomi2009} Tuomi, M., Thum, T., J\"arvinen, H., Fronzek, S., Berg, B., Harmon, M., Trofymow, J. A., Sevanto, S., and Liski, J. 2009. Leaf litter decomposition -- Estimates of global variability based on Yasso07 model. Ecological Modelling, 220, 3362-3371.
\bibitem[\protect\astroncite{Tuomi et al.}{2010a}]{tuomi2010} Tuomi, M., Laiho, R., Repo, A., and Liski, J. 2010. Wood decomposition model for boreal forests. Ecological Modelling, 222, 709-718.
\bibitem[\protect\astroncite{Tuomi et al.}{2010b}]{tuomi2010b} Tuomi, M., Thum, T., and Liski, J. 2010. Modelling soil carbon stock in global scale. In preparation.
\bibitem[\protect\astroncite{Updegraff et al.}{2010}]{updegraff2010} Updegraff, K., Zimmermann, P. R., Kozak, P., Chen, D.-G., and Price, M. 2010. Estimating the uncertainty of modeled carbon sequestration: The GreeCert$^{\textrm{TM}}$ system. Environmental Modelling and Software, 25, 1565-1572.
\bibitem[\protect\astroncite{V\'av\v{r}ov\'a et al.}{2009}]{vavrova2009} V\'av\v{r}ov\'a, P., Penttil\"a, T., and Laiho, R. 2008. Decomposition of Scots pine fine woody debris in boreal conditions: Implications for estimating carbon pools and fluxes. Forest Ecology and Management, 257, 401-412.
\bibitem[\protect\astroncite{Wallman et al.}{2006}]{wallman2006} Wallman, P., Belyazid, S., Svensson, M. G. E., and Svendrup, H. 2006. DECOMP -- a semi-mechanistic model of litter decomposition. Environmental Modelling and Software, 21, 33-44.
\end{thebibliography}
\end{document}